\documentclass[conference]{IEEEtran}

\usepackage{svg}
\usepackage{multirow}
\usepackage{amsmath}
\usepackage{float}
\usepackage{subcaption}
\usepackage{graphicx}
\usepackage[utf8]{inputenc} 
\usepackage[T1]{fontenc}    
\usepackage{hyperref}       
\usepackage{url}            
\usepackage{booktabs}       
\usepackage{amsfonts}       
\usepackage{nicefrac}       
\usepackage{enumitem}

\title{RFSoC based LLRF system design at ALS}

\author{
    \IEEEauthorblockN{
        Qiang Du,
        Shreeharshini Murthy,
        Victoria Moore,
        Angel Jurado Lopez, \\
        Michael Chin,
        Shree Subhasish Basak,
        David Nett,
        Benjamin Flugstad\IEEEauthorrefmark{1}
    }
    \IEEEauthorblockA{
        Lawrence Berkeley National Laboratory\\
        1 Cyclotron Rd, Berkeley, CA, 94720 USA
    }
}

\begin{document}
\maketitle

\begin{abstract}
    The Advanced Light Source (ALS) at LBNL is upgrading several LLRF systems for its
    Linac and Sub-Harmonic Bunchers, where it is desired to have a unified LLRF system
    design to support various RF frequencies (at 125MHz, 500MHz and 3GHz) and configurations.
    This paper demonstrates an open-source, direct sampling RFSoC based LLRF system design,
    featuring: sample-to-sample Multi-Tile Synchronization, deterministic latency,
    digital up/down conversion, arbitrary waveform generation and acquisition,
    in-pulse closed loop control, timing and EPICS integration, modular RF frontend
    and hardware designs.
    Measured RF characteristics show that the RFSoC based LLRF system is able to meet
    the system requirements.
\end{abstract}

\section{Introduction}

The Advanced Light Source (ALS) at Lawrence Berkeley National Laboratory is a
U.S. Department of Energy's synchrotron light source user facility that is
operational since 1993.  With circumference of 196.8 m, the ALS Storage Ring
(SR) keeps electron beam current of 500 mA at 1.9 GeV under multi--bunch mode
user operation to deliver synchrotron X-rays to surrounding 40 experimental end
stations.

Including the on-going upgrade project to the 4th generation light source (ALS-U),
the ALS accelerator complex has developed many digital Low-Level RF (LLRF) control
 systems ~\cite{qiang2017digital}, as shown in Table ~\ref{table:als_llrf} and
 Fig. ~\ref{table:als_llrf}.

\begin{table}[H]
    \centering
    \begin{tabular}{l|c|c|c}
    \toprule
        LLRF System  & Freq. (MHz) & Year & Status     \\
        \midrule
        Storage Ring~\cite{du2019digital}     & 499.64 & 2017 & In operation  \\
        Accumulator Ring~\cite{du2022digital} & 500.39 & 2021 & In commission \\
        Buncher              & 124.91, 499.64 & 2022 & In operation \\
        Electron Gun         & 124.91         & 2023 & In operation \\
        Linac                & 2997.84        & 2025 & In development \\
        Buncher (closed loop)& 124.91, 499.64 & 2025 & In development \\
        Booster Ring         & 499.64         &      & Planed \\
    \bottomrule
    \end{tabular}
    \caption{Roadmap of LLRF systems at ALS in LBNL}
    \label{table:als_llrf}
\end{table}

The Storage Ring LLRF and Accumulator Ring LLRF systems are continuous-wave (CW),
near 500MHz RF frequencies, each consists of 2 normal conductive cavities.
The ALS injection RF systems, including the Electron Gun, Linac and sub-Harmonic bunchers,
requires various RF frequencies and pulse length in pulsed mode.
All LLRF systems are
referencing the master oscillator through the ALS frequency distribution system ~\cite{betz2019},
and subscribe to the ALS timing and event distribution system.
To accommodate the various LLRF requirements and provide a common digital control platform,
we have evaluated and designed a RFSoC based, direct-sampling LLRF system with modular analog frontend.

\begin{figure}[H]
    \centering
    \includegraphics[width=0.75\linewidth]{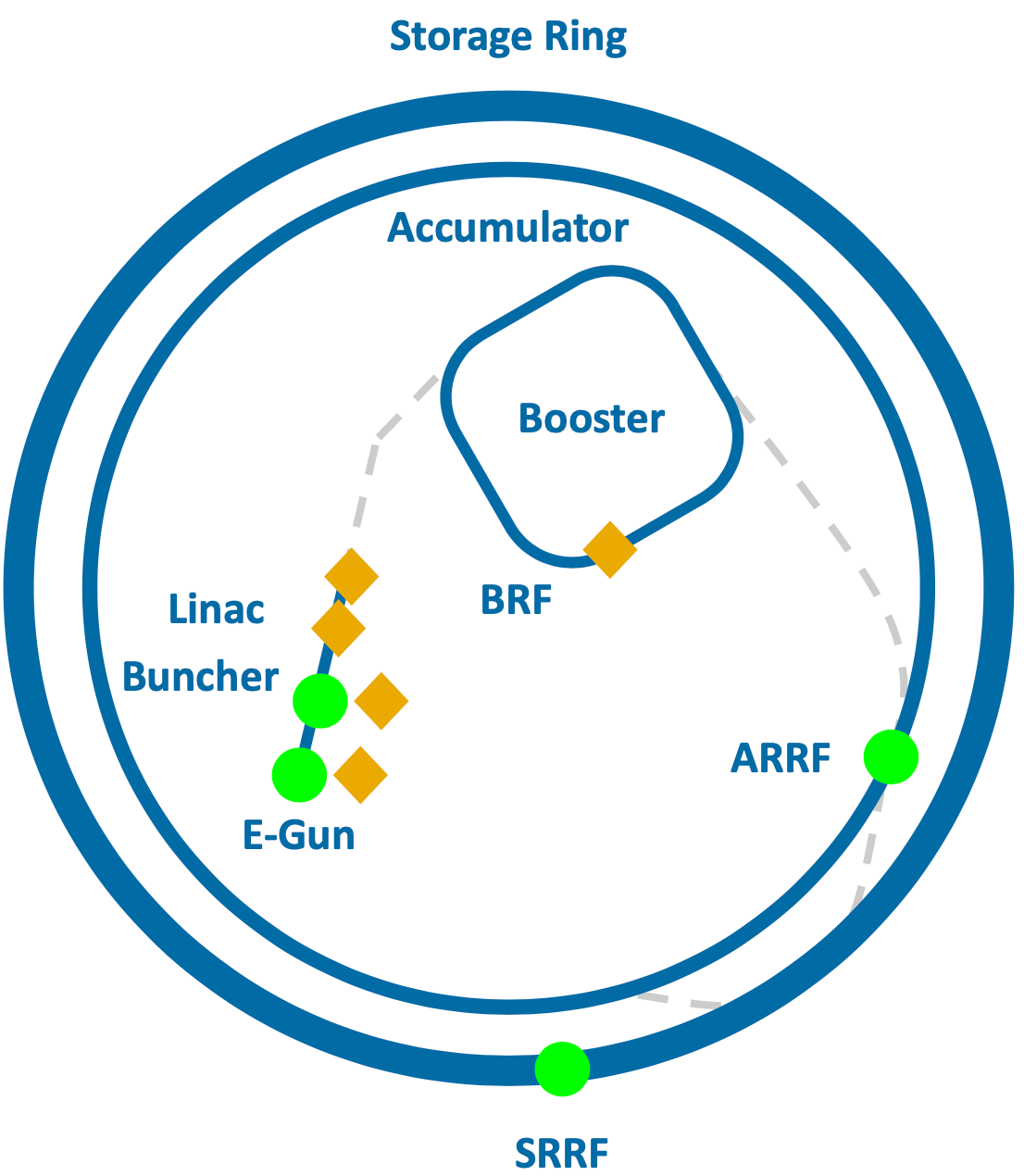}
    \caption{Roadmap of digital LLRF systems at ALS in LBNL.
    Diamond marks planned RFSoC based LLRF systems.}
    \label{fig:als_llrf}
\end{figure}

RFSoC based accelerator instrumentation has been widely adopted for building Beam Position Monitor electronics.
This includes the BPM electronics at ALS-U, LBNL ~\cite{weber:ipac2021-wepab321},
HL--LHC BPMs at CERN ~\cite{irene:2023-cern},
and NSLS--II bunch by bunch BPMs at BNL ~\cite{ha:ipac2024-wepg08}.

For digital LLRF control systems,
there has also been many RFSoC based developments reported,
including the first reported 20 GHz LLRF system in 2019 ~\cite{rohlev201920ghzlownoise},
a recently developed Linac LLRF system at SLAC in 2025 ~\cite{liu:2025compactlowlevelrfcontrol},
and HL--LHC crab cavity LLRF system evaluation in 2025 ~\cite{daniel:2025}.

This paper describes the system design, implementation and preliminary test results for the
 Linac and closed-loop Buncher RFSoC LLRF prototype system at ALS in Berkeley Lab.

\section{System requirements and Overview}

\subsection{Linac LLRF system requirements}

Table ~\ref{table:linac_requirements} describes the RF requirements for
the Linac LLRF system, which consists of two modulators at frequency
of $2997.84$ MHz in pulsed mode.

It is desired to have separate LLRF system each driving one modulator.
Due to the short RF pulse length, inter-pulse RF regulation on amplitude
and phase is required, to maintain $0.1\%$ and $0.1^\circ$ RMS RF field stability.

\begin{table}[H]
    \centering
    \begin{tabular}{l|c|c|c}
    \toprule
        & Description & Value & Unit     \\
    \midrule
        & Num. of RF amplifiers & 2 & \\
        & Num. of LLRF stations & 2 & \\
        & Num. of RF inputs per station & $> 5$ & \\
        & Num. of RF drives per station & $1$ & \\
    \midrule
        $f_\text{MO}$ & MO Frequency            & 499.64  & MHz \\
        $f_\text{RF}$ & $6 \times f_\text{MO} $ & 2997.84 & MHz \\
        $t_\text{pulse}$ & RF Pulse length      & $\sim 5$ & $\mu$s \\
        $f_\text{trig}$ & Trigger rate          & 0.7 & Hz \\
        $f_\text{EVR}$ & Timing EVR clock freq. & $f_\text{MO}/4$ & MHz \\
    \midrule
        & Feed forward & No & \\
        & Feedback mode & inter pulse & \\
        & Feedback loops & amplitude and phase & \\
        & Amplitude stability   & RMS $< 0.1$ & \% \\
        & Phase stability       & RMS $< 0.1$ & $^\circ$ \\
    \bottomrule
    \end{tabular}
    \caption{Linac LLRF requirements}
    \label{table:linac_requirements}
\end{table}

\subsection{Buncher RF system requirements}

Table ~\ref{table:buncher_requirements} describes the RF requirements for
the sub-harmonic buncher LLRF system, which consists of two buncher cavities
at frequencies of $124.91$ and $499.64$ respectively, each driven by a
solid state amplifier (SSA) with 25kW output power.

\begin{table}[H]
    \centering
    \begin{tabular}{l|c|c|c}
    \toprule
        & Description & Value & Unit     \\
    \midrule
        & Num. of RF amplifiers & 2 & \\
        & Num. of LLRF stations & 1 & \\
        & Num. of RF inputs per station & $8$ & \\
        & Num. of RF drives per station & $2$ & \\
    \midrule
        $f_\text{MO}$   & MO Frequency          & 499.64  & MHz \\
        $f_\text{RF1}$  & $f_\text{MO}/4$       & 124.91  & MHz \\
        $f_\text{RF2}$  & $f_\text{MO}$         & 499.64 & MHz \\
        $t_\text{pulse}$ & RF Pulse length      & $\sim 30$ & $\mu$s \\
        $f_\text{trig}$ & Trigger rate          & 0.7 & Hz \\
        $f_\text{EVR}$  & Timing EVR clock freq. & $f_\text{MO}/4$ & MHz \\
    \midrule
        & Feed forward & Yes & \\
        & Feedback mode & intra pulse & \\
        & Feedback loops & amplitude and phase & \\
        & Amplitude stability & RMS $< 0.1$ & \% \\
        & Phase stability & RMS $< 0.02$ & $^\circ$ \\
    \bottomrule
    \end{tabular}
    \caption{Sub-harmonic buncher LLRF requirements}
    \label{table:buncher_requirements}
\end{table}

It is desired to have a single LLRF system driving both SSA stations.
The RF fields are required to be regulated within
the pulse length of $\sim 30 \mu$s, therefore feed-forward control
is likely needed to achieve fast loop settling time, in order to maintain
$0.1\%$ and $0.02^\circ$ RMS RF field stability.

The existing tube based RF High Power Amplifier (HPA) for SHB cavities will 
be replaced and upgraded with solid state LDMOS/ GaN based new 25 kW pulsed 
RF HPA having many advantages and superior performances such as no high \
voltages, modular construction, graceful degradation of RF output power, 
better phase 
noise, high reliability etc. The HPA shall have gating control signal for 
Gate bias for reducing the HPA power dissipation  and the RF input drive 
signal within such gated time from LLRF control. The HPA shall be self 
protecting system will slow and fast interlocks and the HPA can be operated 
in local mode through HMI and remote mode via EPICS interface over 
EtherNet/IP. The RF pulse ON time can be varied From 15 us to 60us and also 
the pulse repetition rate can be varied from 0.5 Hz to 10 Hz. For closed loop 
LLRF control of SHB RF cavities at 125 MHz and 500 MHz with large control 
bandwidth of multi tens of kHz, the HPA  gain shall be more than 74 dB, HPA 
group delay shall be less than 150 ns, the AM to PM conversion shall be less 
than  0.15 deg / dB, RF output power stability (long term) of < 2 \% and RF 
phase stability (long term) of < $4^{\circ}$ etc. The LLRF closed loop PI 
control parameters will be tuned to achieve gain margin of greater than 6 dB 
and phase margin of greater than $45^{\circ}$.

\subsection{RFSoC LLRF system architecture}

In order to minimize the engineering and maintenance cost of the digital LLRF systems,
it is desired to have a common platform to accommodate all requirements from both Linac
and sub-harmonic buncher RF systems.

\subsubsection{LLRF Requirements}
\label{sec:llrf_requirements}

Compared with other applications of RFSoC based systems, the digital LLRF system has stringent requirements of:

\begin{description}
    \item[Low TX phase noise:] Additive DAC RMS phase noise jitter to be $<100$ fs [1Hz, 1MHz];
    \item[Low latency:] Feedback control bandwidth requires similar or lower end-to-end latency to be $<300$ ns;
    \item[Low Crosstalk:] Precise control requires accurate measurement of RF signal against crosstalk, to be $>75$ dB;
    \item[Low ADC noise:] Comparable to conventional ADCs with noise spectrum density $<-150$ dBFS/Hz, ENOB of about 9.0 bits;
    \item[Alignment:] Sample-to-sample alignment across all ADC and DAC channels is required;
    \item[Deterministic:] Power-cycle deterministic latency and alignment across all channels;
\end{description}

\subsubsection{System architecture design}

\begin{figure}
    \centering
    \begin{subfigure}[b]{\linewidth}
        \centering
        \includegraphics[width=\linewidth, page=1]{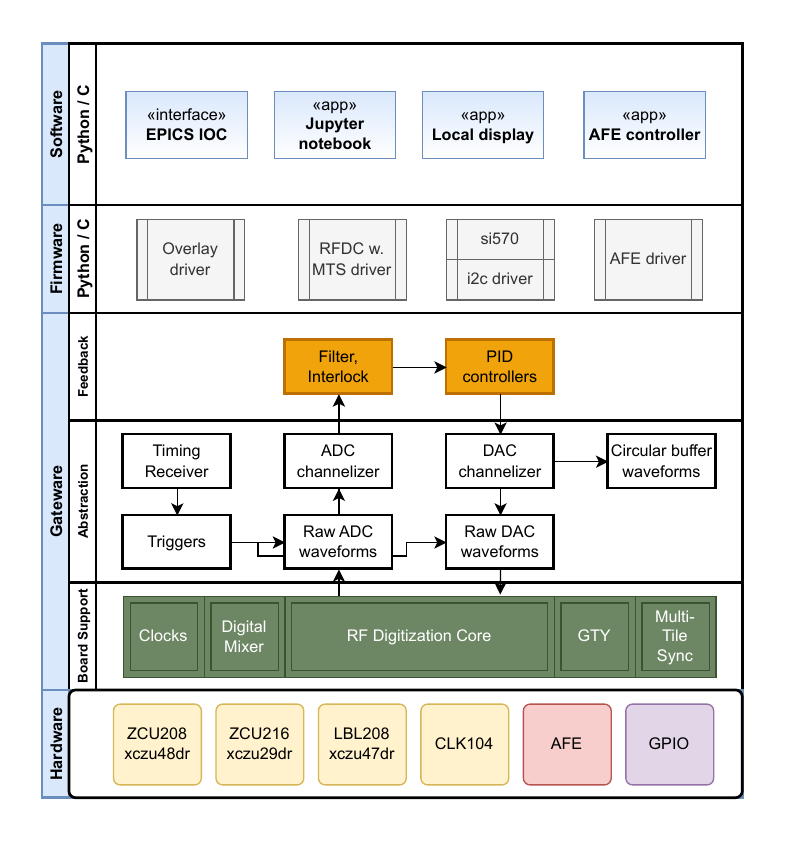}
        \caption{System hardware, firmware and software architecture}
        \label{fig:architecture}
    \end{subfigure}
    \begin{subfigure}[b]{\linewidth}
        \centering
        \includegraphics[width=\linewidth, page=3]{fig/rfsoc_llrf.pdf}
        \caption{System logical view}
        \label{fig:logical_view}
    \end{subfigure}
    \caption{RFSoC LLRF system architecture}
\end{figure}

We designed the RFSoC based LLRF system for ALS following the requirements.
As shown in Fig. ~\ref{fig:architecture}, our RFSoC
based LLRF system is consisted of:

\begin{description}
    \item [Clocking:]
    Generate sampling clocks and reference clocks required by RFSoC. Optimal additive phase noise
    is achieved by re-designing the \href{https://docs.amd.com/r/en-US/ug1437-clk104}{CLK104} board.
    In this design, we choose $f_s = 4$ GHz sampling clock for both ADC and DAC. $f_s$ is generated
    by the LMK04828 and LMX2594 on the CLK104 board with optimal
     configurations using the external reference at $f_\text{MO}$.
    \item [RFSoC board:] Based on
    \href{https://www.amd.com/en/products/adaptive-socs-and-fpgas/evaluation-boards/zcu208.html}{ZCU208} evaluation board,
    we designed a customized board named LBL208 with a lower-end RFSoC chip (\texttt{XCZU47DR-1}).
    The firmware and software architecture also supports ZCU208 and
    \href{https://www.amd.com/en/products/adaptive-socs-and-fpgas/evaluation-boards/zcu216.html}{ZCU216} boards,
    by reusing the majority of DSP RTL implementation and device support layers.
    \item [Analog frontend:] Configurable, modular RF front end platform enables
    adaptability and design reuse. Flexible RF signal conditioning and filtering
    is implemented at a dual-channel slice that is plugged on an interposer board.
    High RF channel to channel isolation is achieved through enclosure shielding and dedicated power rails for each channel.
    \item [Chassis:] A unified, modular and compact chassis design ensures thermal
    and mechanical reliability. Local display and human interface is available for quick
    health diagnostics. Trigger, SFP+, GPIO signals are available through the panel.
    Thermal management is implemented via PMBus by the RFSoC controller.
    \item [Firmware:] An integrated board support layer enables peripheral hardware control,
    including CLK104, analog front-end board, local display and chassis management. The RF digitization core,
    GTY transceivers, Gigabit Ethernet are also integrated via Linux drivers and system booting and initiliazation
    procedure. Digitized RF signals are frequency-converted to base-band for processing, where the firmware design
    section will have detailed description.
    \item [Software:] EPICS IOC runs on-chip at RFSoC board. The multiprocessor Linux running
    on the RFSoC fully takes advantage of the 4 Cortex-A53 cores (at 1.5GHz clock) and has
     direct memory access to PL cores and resources.
    Jupyter notebook and EPICS OPI enables remote development and operations.
\end{description}

\section{System implementation and characteristics}

\subsection{RF performance evaluation}

In-depth comparative evaluation of the LBL208 board is reported using the same firmware design
~\cite{shree:2025}, where conclusions are made:

\subsubsection{DAC phase noise}

It is found that the additive RMS phase jitter from the default ZCU208 kit is about 400 fs [1Hz, 1MHz],
where the majority of the phase noise contribution is from CLK104 board. It is critical to have hardware
optimization of the CLK104 board to optimize the PLL performances in order to achieve the following results.

\begin{figure}[t]
    \centering
        \begin{subfigure}[b]{\linewidth}
        \centering
        \includegraphics[width=.9\linewidth]{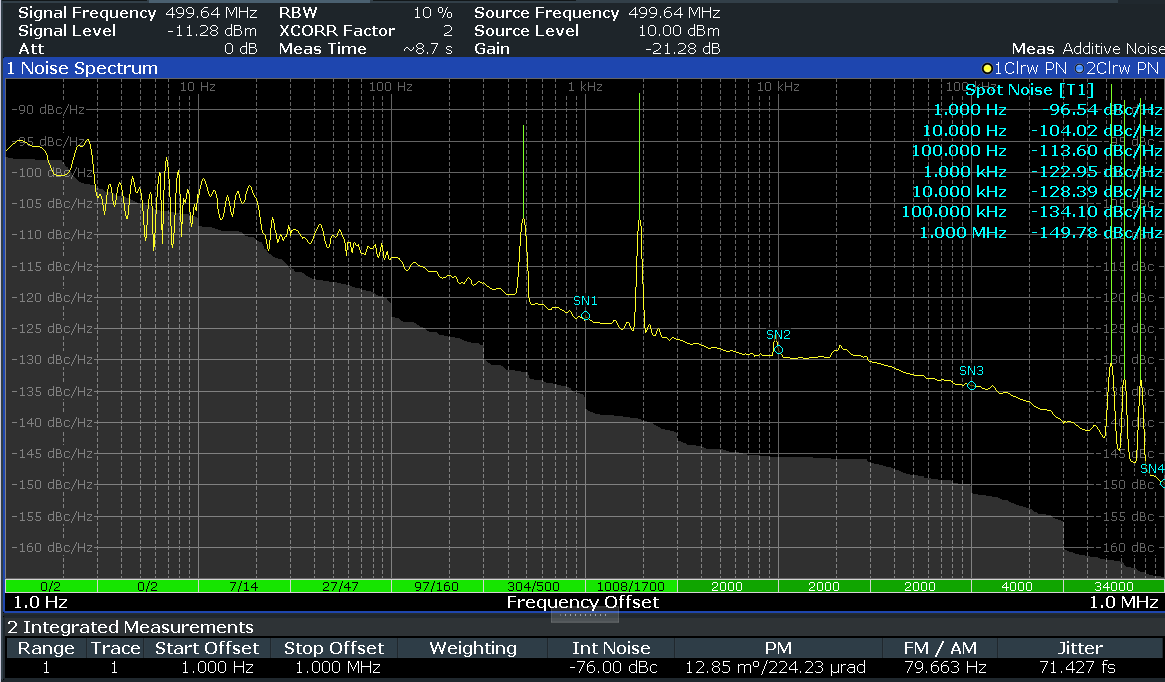}
        \caption{Additive phase noise at 499.64 MHz , RMS jitter is 71.4 fs}
        \label{fig:dac_500MHz_fswp}
    \end{subfigure}
    \begin{subfigure}[b]{\linewidth}
        \centering
        \includegraphics[width=.9\linewidth]{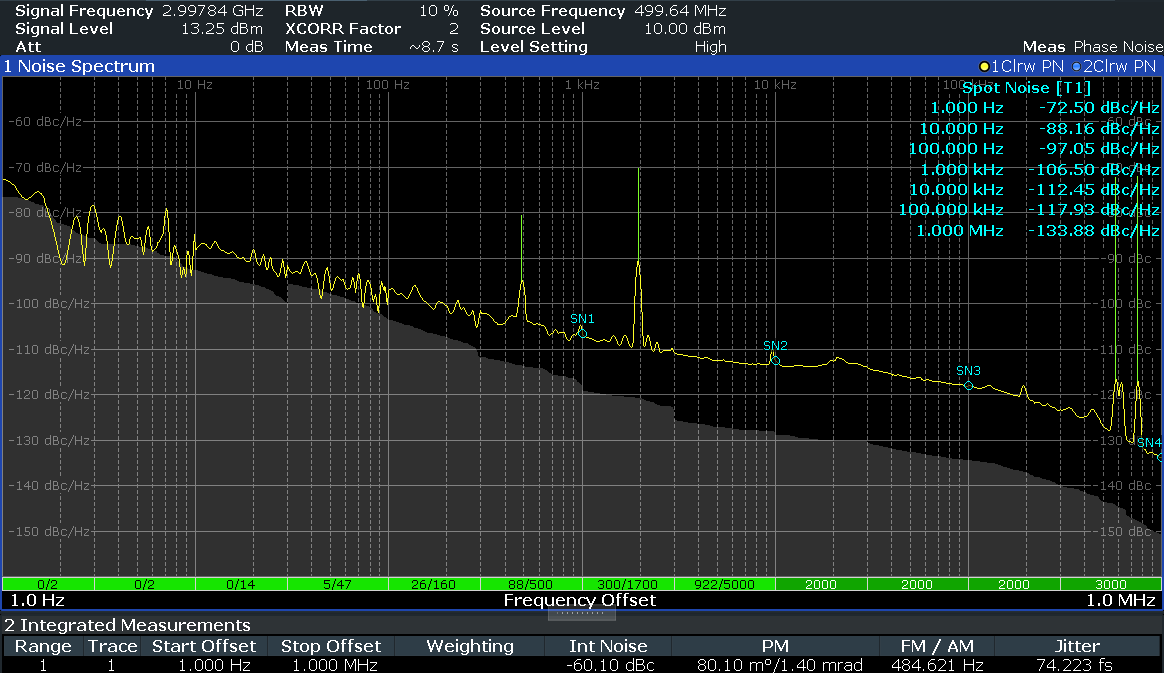}
        \caption{Absolute phase noise at 2997.84 MHz, RMS jitter is 72.8 fs}
        \label{fig:dac_3000MHz_fswp}
    \end{subfigure}
    \caption{RF output phase noise measurements with RMS jitter integrated over [1Hz, 1MHz]}
\end{figure}

Additive phase noise is measured as RMS jitter of $< 80$ fs [1Hz, 1MHz], at 499.64 MHz RF output;
The measurement is done using a R\&S FSWP phase noise analyzer to provide a 499.64 MHz ($f_\text{MO}$)
as the reference to the modified CLK104 board, and directly generate a CW tone at the same frequency.
The results is shown in Fig. ~\ref{fig:dac_500MHz_fswp}.

The absolute phase noise measurement is also conducted in a similar way except the RF output
frequency is set to 2997.8 MHz ($f_\text{RF}$ in ALS Linac), as shown in Fig. \ref{fig:dac_3000MHz_fswp}.

Both cases shows $< 80$ fs [1Hz, 1MHz] RMS phase jitter.
This is a bit worse than the $\sim 30\,$ fs absolute phase jitter measured at the ALS-U LLRF system
 ~\cite{du2022digital} under the same condition.

We estimate that about 60\% of the total phase noise
is from the CLK104 PLL which generates the DAC sampling clock.
This estimate is consistent after comparing with a comprehensive RFSoC
 characteristics reports from AMD ~\cite{amd:2021:rfdac}.

\subsubsection{DAC output spectrum}

\begin{figure}
    \centering
    \includegraphics[width=.9\linewidth, clip, trim=0cm 1.6cm 0cm 2cm]{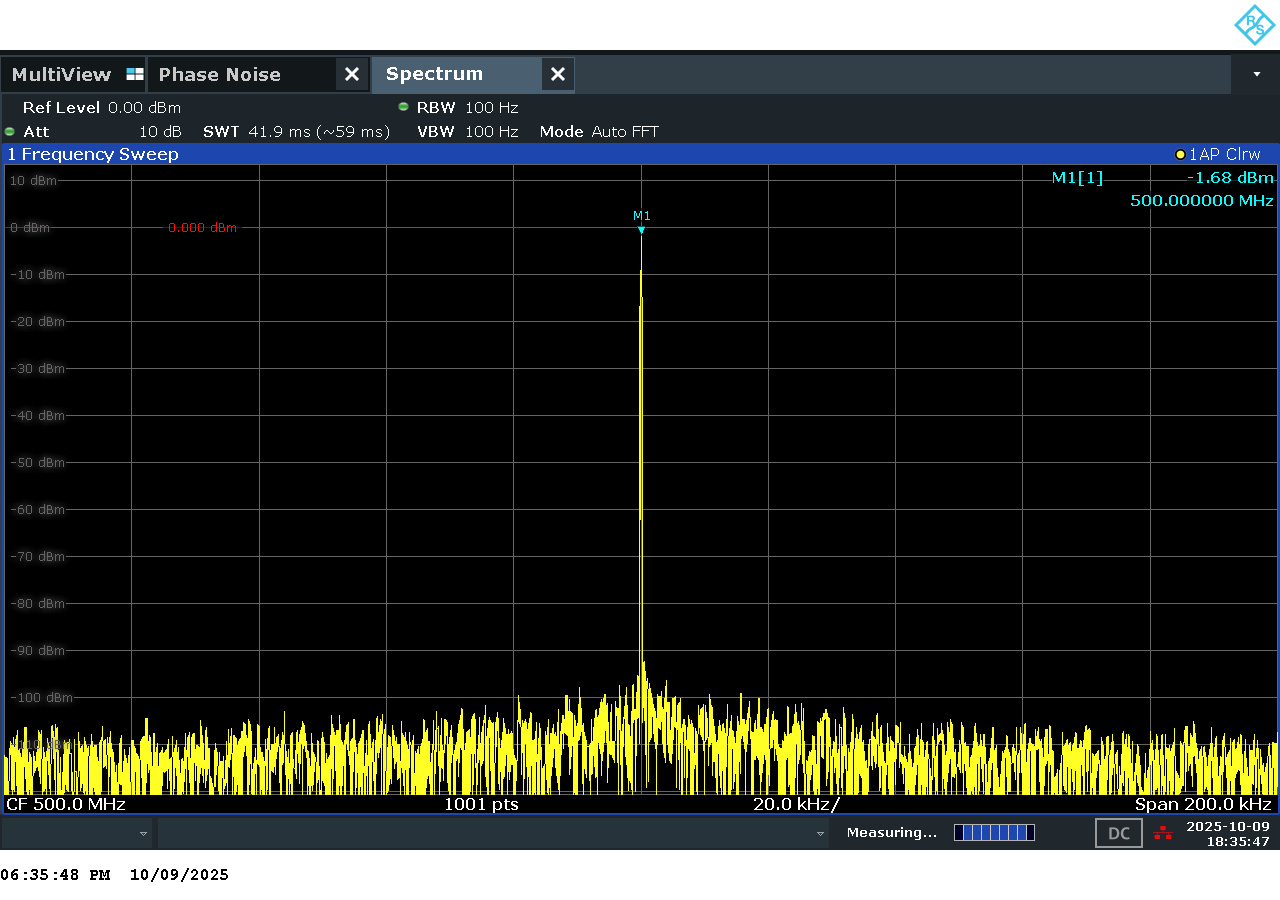}
    \caption{DAC output spectrum at 500 MHz, SFDR is about 90 dB within 200kHz span, 100Hz RBW}
    \label{fig:dac_spectrum}
\end{figure}

Fig ~\ref{fig:dac_spectrum} shows the measured DAC output at 500MHz at narrow band.
The SFDR is similar to the ALS-U LLRF performance~\cite{du2022digital},
which is measured under the same condition.

\subsubsection{ADC characteristics}

By driving the ADCs by a known 500MHz signal by R\&S SMA100B through the XM655,
we measured the ADC SNR of 55.78 dB, SFDR of 79.48 dBc, NSD
of -156.1 dBFS/Hz, and ENOB of 8.97 bits ~\cite{shree:2025}.

\subsubsection{ADC crosstalk}

We measured the crosstalk ADC characteristics by injecting a known 500MHz signal to
 one channel at a time through XM655, and concluded that the worst case crosstalk between any ADC channels
  is better than 75.2\,dB, as shown in Table ~\ref{table:adc_xtalk}.
This is consistent with AMD's \href{https://docs.amd.com/r/en-US/ds926-zynq-ultrascale-plus-rfsoc}{data sheet}
 and RFADC characteristics report~\cite{amd:2021:rfadc}
 which claims better than 75\,dB crosstalk,
 as well as the evaluation benchmarks from CERN for crab cavity LLRF ~\cite{daniel:2025}
 and HL--HLC BPMs ~\cite{irene:2023-cern}.

\begin{table}[H]
\begin{tabular}{|c|r|r|r|r|r|r|r|r|}
\hline
& C0    & C1    & C2    & C3    & C4    & C5    & C6    & C7    \\ \hline
\hline
C0   & -0.4  & -97.4 & -80.5 & -76.5 & -78.9 & -83.3 & -83.4 & -85.9 \\ \hline
C1   & -81.2 & -0.4  & -83.2 & -76.7 & -82.9 & -83.4 & -85.7 & -81.6 \\ \hline
C2   & -82.5 & -88.1 & -0.3  & -75.2 & -82.7 & -83.1 & -86.9 & -86.4 \\ \hline
C3   & -85.7 & -90.9 & -88.4 & -0.0  & -79.9 & -82.6 & -91.2 & -80.8 \\ \hline
C4   & -88.1 & -91.8 & -83.1 & -75.3 & -0.3  & -82.9 & -91.5 & -84.1 \\ \hline
C5   & -86.9 & -92.4 & -82.7 & -76.7 & -84.9 & -0.3  & -84.3 & -81.5 \\ \hline
C6   & -80.9 & -93.3 & -80.2 & -76.4 & -83.7 & -85.1 & -0.3  & -83.0 \\ \hline
C7   & -84.5 & -89.7 & -81.7 & -77.2 & -82.9 & -83.2 & -82.9 & -0.5  \\ \hline
\end{tabular}
\caption{Measured ADC crosstalk at 500MHz input (dB)}
\label{table:adc_xtalk}
\end{table}

Overall the ADC characteristics and crosstalk are similar to a conventional LLRF system
 like the reported ALS-U LLRF performance ~\cite{du2022digital}, and meets the requirements
 listed in Section ~\ref{sec:llrf_requirements}.

\subsection{Hardware Design}

\subsubsection{Analog Frontend}

\begin{figure}
    \centering
    \begin{subfigure}[b]{\linewidth}
        \centering
        \includegraphics[width=.9\linewidth]{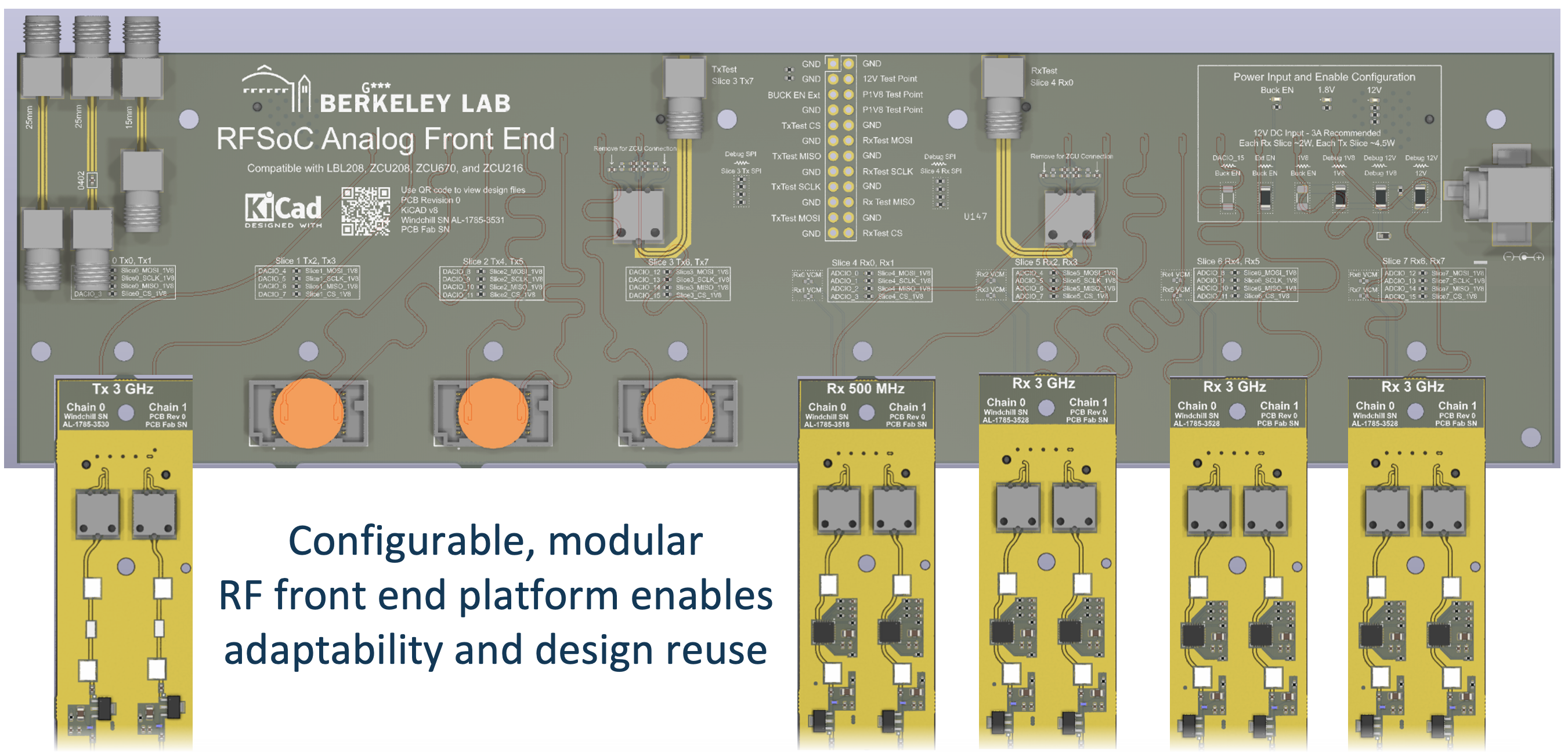}
        \caption{Interposer board}
        \label{fig:frontend}
    \end{subfigure}
    \begin{subfigure}[b]{\linewidth}
        \centering
        \includegraphics[width=.4\linewidth]{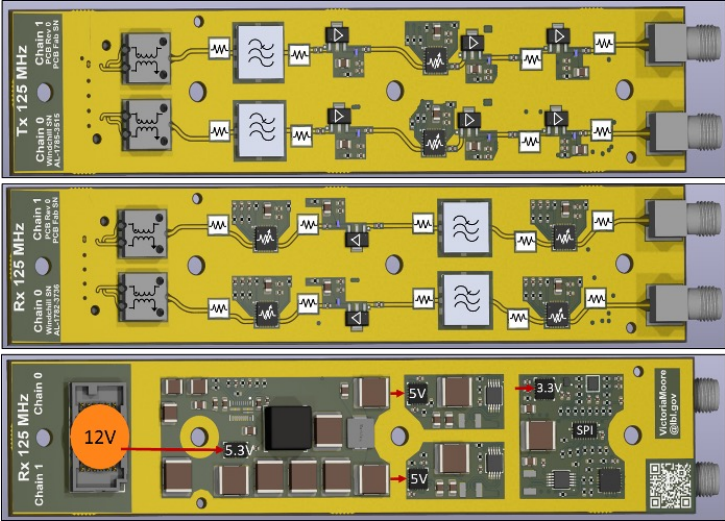}
        \caption{RF slices}
        \label{fig:rf_slices}
    \end{subfigure}
    \caption{Modular Analog Front-End design}
\end{figure}

As shown in Fig ~\ref{fig:frontend}, a passive RF interposer board breaks out the RFSoC board differential RF signals to 4 TX and 4 RX
pluggable modules, or RF slices (shown in Fig ~\ref{fig:rf_slices}), each consisting of 2 channels of receiving or transmitting
RF signal processing chain, with differential to single-ended conversion, RF gain, programmable attenuation, and low-pass filtering
with enclosure shielding for optimal RF isolation performance. The 12V power input from the LBL208 is passed through the interposer,
and local power rails are generated on the slices for each RF channel. Each RF slice expands 4 RFSoC HDIOs into a dedicated SPI bus
for gain control and telemetry from each slice.

\begin{figure}
    \centering
    \begin{subfigure}[b]{.45\linewidth}
        \centering
        \includegraphics[width=.9\linewidth]{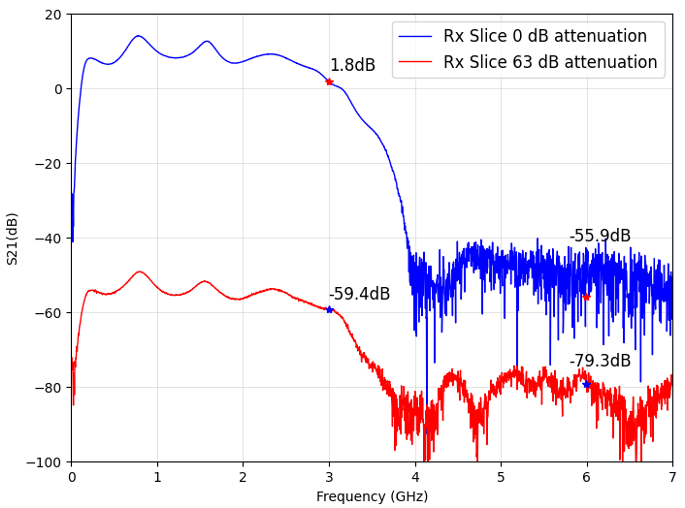}
        \caption{RX Slices Gain}
    \end{subfigure}
    \begin{subfigure}[b]{.45\linewidth}
        \centering
        \includegraphics[width=.9\linewidth]{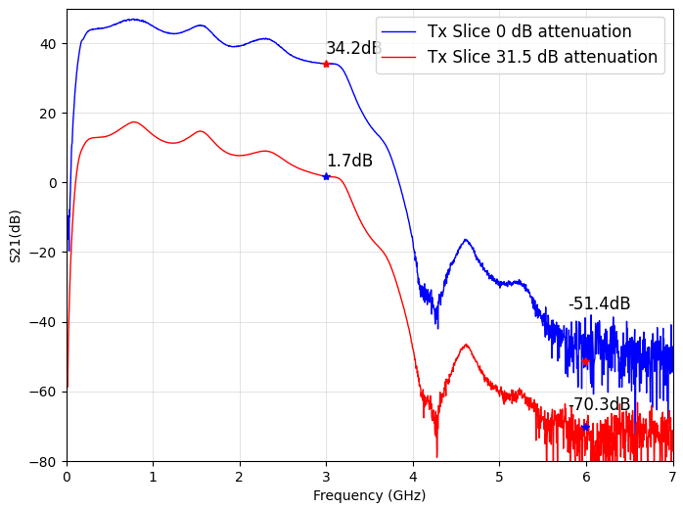}
        \caption{TX Slices Gain}
    \end{subfigure}
    \begin{subfigure}[b]{.45\linewidth}
        \centering
        \includegraphics[width=.9\linewidth]{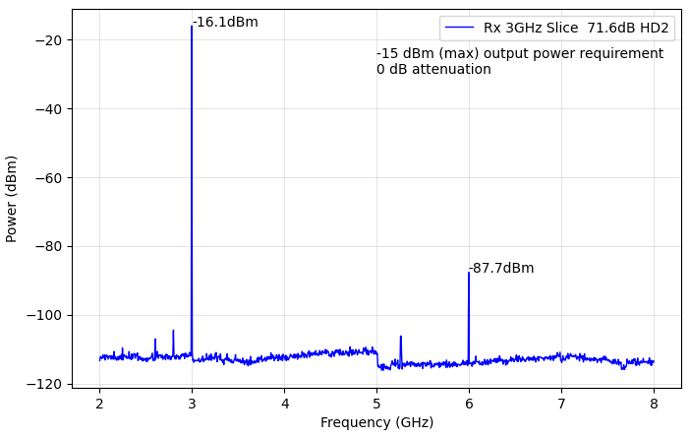}
        \caption{RX Slices Linearity}
    \end{subfigure}
    \begin{subfigure}[b]{.45\linewidth}
        \centering
        \includegraphics[width=.9\linewidth]{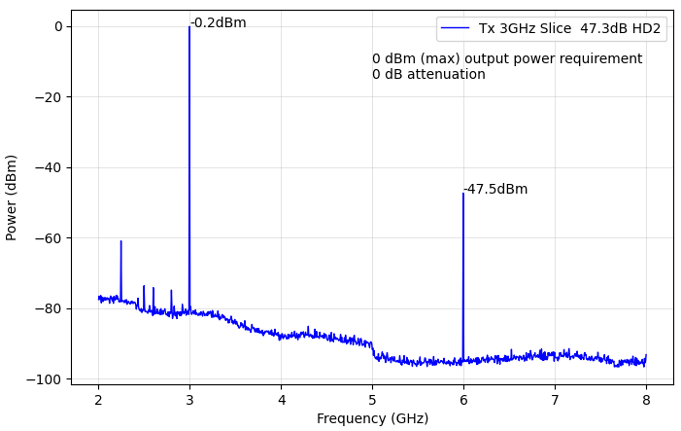}
        \caption{TX Slices Linearity}
    \end{subfigure}
    \begin{subfigure}[b]{.9\linewidth}
        \centering
        \includegraphics[width=.5\linewidth]{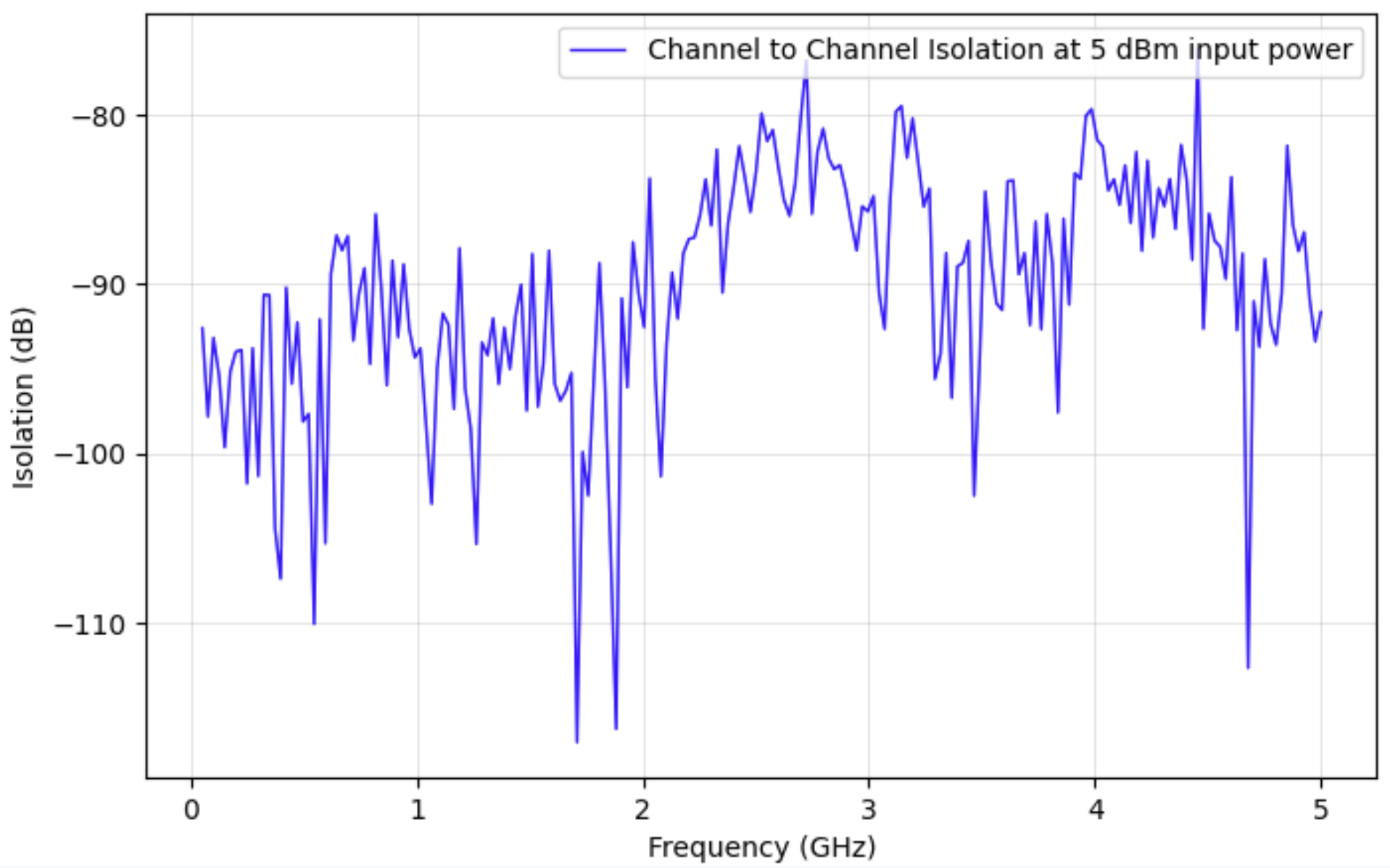}
        \caption{Rx Slice inter-channel isolation}
    \end{subfigure}
    \caption{RF slices characteristics}
    \label{fig:rf_slices:characteristics}
\end{figure}

As shown in Fig ~\ref{fig:rf_slices:characteristics}, the measured RF performances of
 RF slices are:

\begin{description}
    \item[RX gain:]
    63 dB dynamic range, 0.5 dB resolution, second harmonic 20--50 dBc.
    \item[TX gain:]
    31.5 dB dynamic range, 0.5 dB resolution, second harmonic 70 dBc.
    \item [RX linearity:]
    second harmonic is 71.6 dB lower than carrier at 3GHz, which is $>20$ dB better than ADC linearity.
    \item [TX linearity:]
    second harmonic is 47.3 dB lower than carrier at 3GHz, which is 7 dB better than pre-amplifier.
    \item[RX additive phase jitter:] 0.41 fs [1Hz, 1MHz] at 3GHz.
    \item[TX additive phase jitter:] 1.09 fs [1Hz, 1MHz] at 3GHz.
    \item[RX channel to channel (within a Slice) isolation:] >75 dB [100MHz, 5GHz]
\end{description}

\begin{figure}
    \centering
    \begin{subfigure}[b]{0.45\linewidth}
        \centering
        \includegraphics[width=\linewidth, clip, trim=2.5cm 2.5cm 17.5cm .8cm]{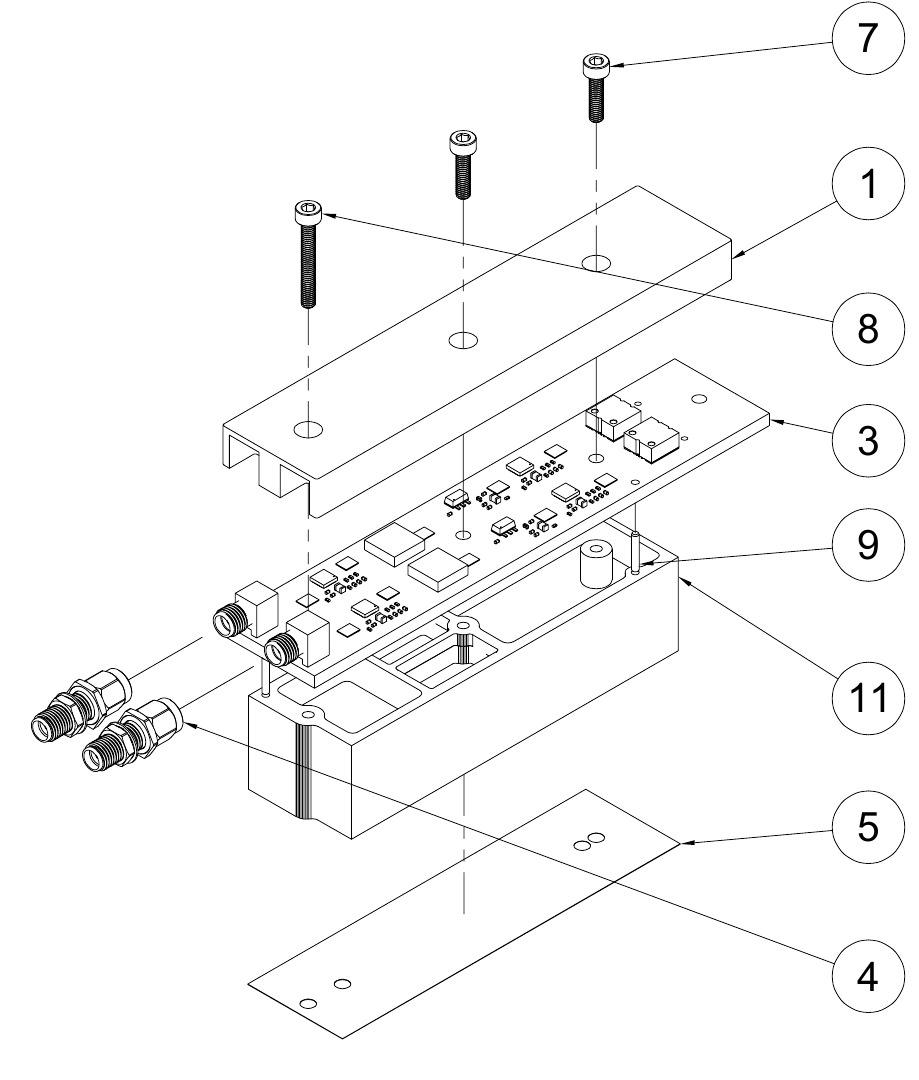}
        \caption{RF slices PCB and shielding assembly}
    \end{subfigure}
    \hfill
    \begin{subfigure}[b]{0.45\linewidth}
        \centering
        \includegraphics[width=\linewidth, clip, trim=4.5cm 8cm 22.5cm 4cm]{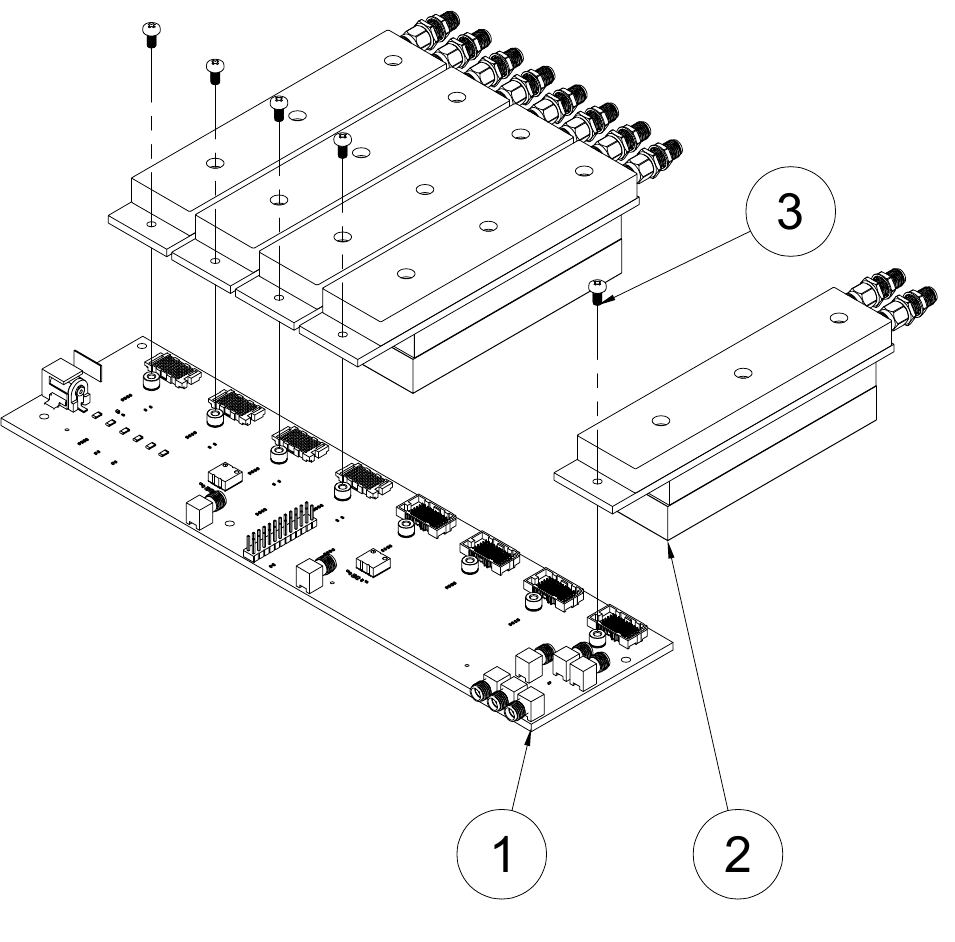}
        \caption{Interposer and RF slices assembly}
    \end{subfigure}
    \caption{RF analog frontend assembly}
    \label{fig:zfff_assy}
\end{figure}

Fig ~\ref{fig:zfff_assy} shows the shielding and mechanical assembly of the RF slices,
and the mounting assembly of them on the interposer board.

\subsubsection{Enclosure}

\begin{figure}[H]
    \centering
    \includegraphics[width=.8\linewidth]{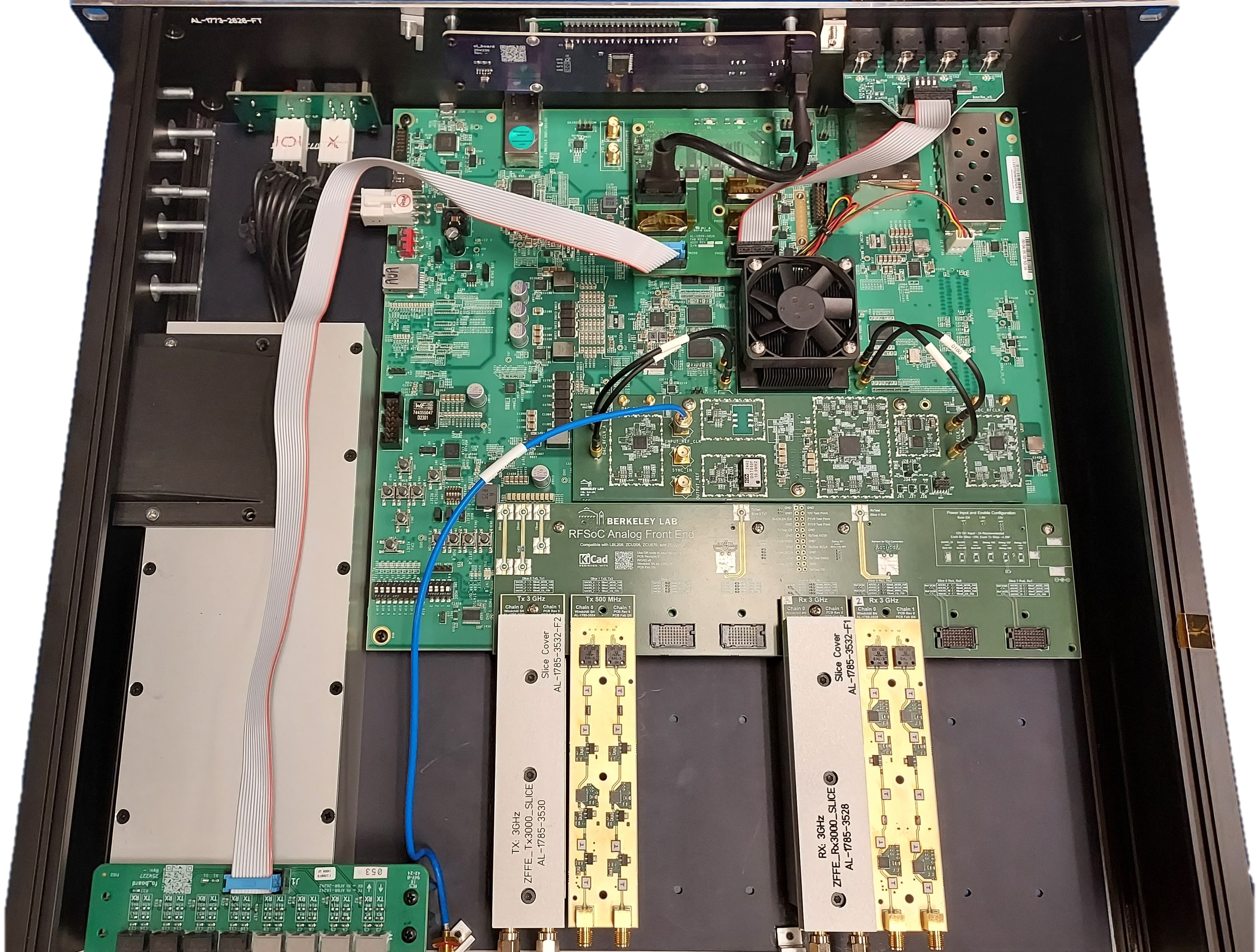}
    \caption{Assembled LLRF chassis}
    \label{fig:chassis}
\end{figure}

Putting things together,
Fig ~\ref{fig:chassis} shows the assembly of the chassis mechanical
design, which allows rigid, cable-less RF connections with integrated passive thermal management.
 A pluggable computer power supply
is included in the chassis with PMBus management interface connected to the RFSoC for health monitoring
and self-protection against unexpected hardware failure or environmental conditions.

\subsection{Firmware Design}

\begin{figure}
    \centering
    \includegraphics[width=\linewidth,page=1]{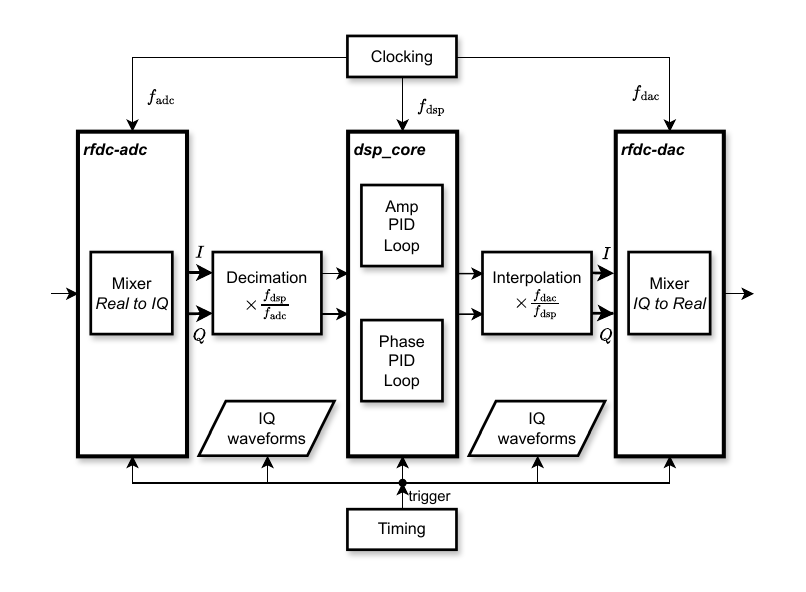}
    \caption{Firmware framework and signal flow}
    \label{fig:dsp_flow}
\end{figure}

As part of the system architecture as shown in Fig ~\ref{fig:architecture},
the LLRF firmware and signal flow is illustrated in Fig ~\ref{fig:dsp_flow}.

As part of the board support design,
the design used AMD's RFSoC RF Data Converter (RFDC) IP core as specified in PG269 ~\cite{amd:pg269},
together with other IP cores including the Zynq MPSoC Processing System
 (\href{https://docs.amd.com/r/en-US/pg201-zynq-ultrascale-plus-processing-system}{PG201}),
 GTY transceiver wizard
  (\href{https://www.xilinx.com/support/documents/ip_documentation/gtwizard_ultrascale/v1_7/pg182-gtwizard-ultrascale.pdf}{PG182}),
  and others.

  The firmware design ~\cite{lbnl:pynq_llrf} also utilized common RTL modules from LBNL
 \href{https://github.com/berkeleylab/bedrock}{Bedrock} repository
 and the USPAS LLRF repository ~\cite{lbnl:uspas_llrf}.

All RTL designs are tested using \href{https://www.cocotb.org/}{cocotb} for behavioral validation.
Common RTL and IP cores are put together by automated scripts to support the applications
 of narrow-band LLRF and wid-band MIMO control on
3 different RFSoC platforms:
\begin{description}
    \item [AMD ZCU208:] \verb|xczu48dr-2fsvg1517e|
    \item [AMD ZCU216:] \verb|xczu29dr-2ffvf1760e|
    \item [LBNL LBL208:] \verb|xczu47dr-1fsvg1517e|
\end{description}

\subsubsection{Clocking}

As shown in Fig ~\ref{fig:logical_view}, there are a total of 4 clocks generated from an external reference
$f_\text{MO}$, including ADC and DAC sampling clocks, system reference clock $f_\text{SYSREF}$,
 and the programmable logic clock $f_\text{PL}$.
The LLRF digital signal processing clock is derived from $f_\text{PL}$ by $f_\text{DSP} = f_\text{PL} / 2$.
The clock generation and synchronization steps follows the requirements for
 \href{https://docs.amd.com/r/en-US/pg269-rf-data-converter}{Multi-Tile Synchronization (MTS)}.

Given $f_\text{MO}=500$ MHz, the design currently have the same sampling frequencies at 4000 MHz for both ADC and DAC.
The sampling frequencies are generated by external LMX2594 chips on the CLK104 board, because this allows PLL performance
tunning and lower additive phase noise compared to the on-chip PLLs at the RFSoC.

\subsubsection{Sampling and freuency conversion}

Thanks to the digital mixers provided in the RFDC core, the firmware is able to convert the digitized signal
to base-band using the real to complex fine mixer setting in the RF-ADC,
and from base-band using the complex to real fine mixer setting in the RF-DAC.
The RFDC mixers allows various sampling schemes including odd or even Nyquist zone.

In particular, we use the second Nyquist zone sampling for generating RF frequency at 3GHz for ALS Linac LLRF system,
and the measured spectrum is shown in Fig ~\ref{fig:dac_3000MHz_fswp}.
Similar to discrete RF-DAC chips, the RF-DAC has mix-mode to compensate the frequency response in the second Nyquist Zone.
We have compared the performance against a 7GHz sampling scheme (highest allowable with MTS, signal is at first Nyquist Zone),
and concluded that there was no significant improvement in terms RF power or SNR by the higher sampling rate.

All mixer parameters, including frequency, Nyquist zone, phase, etc., are made run-time configurable by software.
This allows quick troubleshooting and development cycles.

Non-IQ sampling can be easily achieved by choosing a different ratio between the RF and sampling clock frequencies,
and we have experimented and confirmed its functionality but not fully characterized the RF performances.

\begin{figure}
    \centering
    \includegraphics[width=\linewidth,page=2]{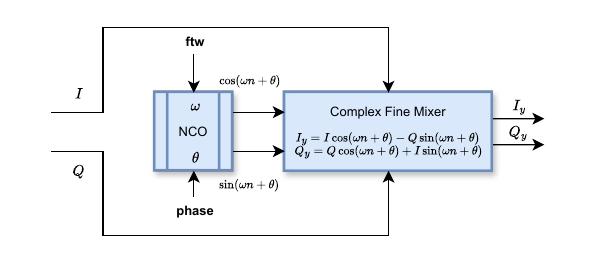}
    \caption{Digital RF mixers for RF-ADC and RF-DAC}
    \label{fig:dsp_mixer}
\end{figure}

\subsubsection{Data alignment}

Based on Multi-Tile Synchronization (MTS), a total of 3 layers of data alignments are implemented
 to satisfy the system requirements ~\ref{sec:llrf_requirements}:
\begin{description}
    \item[MTS channel alignment:] The same relative latencies across all ADC and DAC channels are guaranteed upon system reconfiguration and power cycle;
    This involves measurement and maintaining of clock phases inside the RFDC digital signal path across clock domains, using $f_\text{SYSREF}$.
    \item[MTS deterministic latency:] The end-to-end latency, across all ADC and DAC channels are guaranteed upon system reconfiguration and power cycle;
    \item[Mixer NCO phases:] The relative phase of each digital mixers in RFDC ADC and DAC is reset at the same time, to guarantee the generated
     RF signal's phase relative to external reference;
\end{description}

The design follows procedures described in PG269 ~\cite{amd:pg269}, and tested by a full loopback between each DAC and ADC channel.

\subsubsection{Timing interface}

\href{http://www.mrf.fi/index.php/timing-system}{MRF Timing System} compatible event receiver (EVR) is implemented
by a RTL design together with GTY transceiver wizard IP core. A specific event code can be subscribed through
 the deserialized 8B10B data stream from the EVR interface,
at the time resolution of about 8 ns.
The power-cycle deterministic timing of recovered event is guaranteed and tested against
the global MRF event generator (EVG).
A watch-dog logic is designed to reset and recover the EVR status, in case of interruption of the timing stream (e.g. re-plug of fiber-optic cable).

\subsubsection{LLRF controllers}

\begin{figure}[h]
    \centering
    \includegraphics[width=\linewidth,page=3]{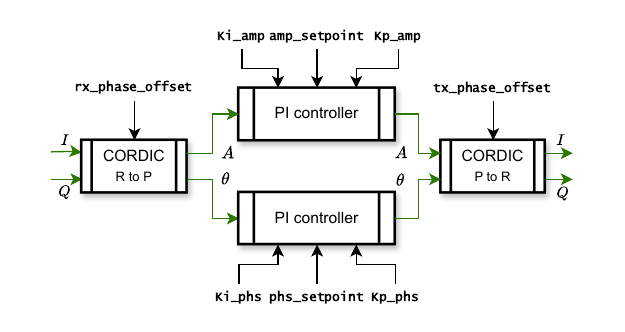}
    \caption{Loop Controllers}
    \label{fig:dsp_core}
\end{figure}

LLRF DSP and controllers are identical to the existing LBNL design from the USPAS LLRF repository ~\cite{lbnl:uspas_llrf},
where a complex pair of measured base-band RF signals are converted to amplitude and phase data streams,
and are regulated by independent PID feedback controller, before the control output being converted back to
the complex RF pair for up-conversion. The loop gain and enable controls are available through each PID controller's register
 interface. Fig ~\ref{fig:dsp_core} shows the LLRF loop controllers block diagram.

Wide-band raw waveforms, or base-band IQ waveforms are available using on-chip ultra-ram resources.
Additionally, the RFSoC boards also have 8G bytes of DDR4 memory for deep local storage.

\subsubsection{Intra-pulse feedback}

\begin{figure}[h]
    \centering
    \includegraphics[width=\linewidth]{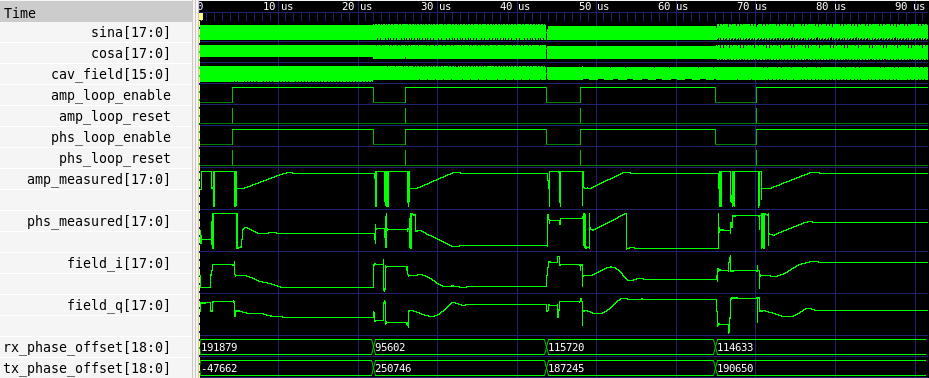}
    \caption{Simulation of feedback loops}
    \label{fig:uspas_simulation}
\end{figure}

Closed loop feedback control is numerically simulated and validated against cavity parameters, to take account of total latency.
Fig ~\ref{fig:uspas_simulation} shows the dynamics of the loop closing using the same controller ~\cite{lbnl:uspas_llrf} in base-band simulation.
By opening and closing the loop (4 times) at random set-points, it shows the settling time of about $10\mu $s can be achieved.
Based on the measured end-to-end latency of $< 300$ns, it is estimated that the intra-loop feedback
 control of the ALS Linac is feasible within $30 \mu$ s pulse length.

\subsection{Software Design}

\subsubsection{Architecture}
The LLRF system is fully leveraging the quad Cortex-A53 core (at 1.5GHz clock frequency) by running
on a petalinux (based on Ubuntu 22.04). The Linux device tree and drivers made peripheral management easily
integrated with the firmware design. We choose the \href{https://www.pynq.io/}{PYNQ} framework, where
a pre-built linux image is suitable and customized to run the firmware overlay, which is consisted by the bitstream
 file and the hardware description file for register maps.
IP cores and DSP registers are integrated by encapsulating the designs in a modular, block design process.
The PYNQ framework also provides Jupyter notebook and Jupyter labs running on the chip, making the testing
 and troubleshooting process easy and tool-free.

\subsubsection{Overlay driver}
A Python Overlay driver class (\verb|mimo_mts|) is provided with configurable options
for supporting different configurations
of RFDC, clocking, mixers and LLRF DSPs through their registers and software drivers.
Sub classes of different LLRF and MIMO applications are implemented for each application.
The configurations are initialized at boot time and can be re-configured at run-time.

\subsubsection{Peripheral support}
At boot time, the system configures peripherals through Linux device drivers. For example, the LMK04828 and LMX2594 chips
on the CLK104 boards are programmed using the text-based register files generated by TI's
 \href{https://www.ti.com/tool/TICSPRO-SW}{TICS Pro} software. This allows
easy management of many flavors of overlays, over many RFSoC platforms supported by the same repository ~\cite{lbnl:pynq_llrf}.
Similarly, the DDR4 PL memory, GPIOs and other peripherals are dynamically loaded by the overlay Python driver at boot time
 through the Linux device tree system.

The enclosure power and thermal management is also done through the Linux device driver, via the PMBus and standard software tools.

\subsubsection{EPICS IOC}

Thanks to the Python packaging and modular interface to the PYNQ overlay driver, it is easy to implement a
 \href{https://github.com/DiamondLightSource/pythonSoftIOC}{pythonSoftIOC} that runs on the petalinux.
 The register addresses and waveforms are directly mapped to the Linux memory space, therefore the high reading and writing speed allows
 efficient updating the EPICS PVs through a polling loop.

 For example, all IQ waveforms (64k samples each) of each ADC and DAC channels are also converted to amplitude and phase waveforms.
 It is tested that at a polling rate of 10 Hz, the CPU load is 60\%, out of total of 400\%.

\begin{figure}[h]
    \centering
    \begin{subfigure}[b]{0.48\linewidth}
        \centering
        \includegraphics[width=\linewidth]{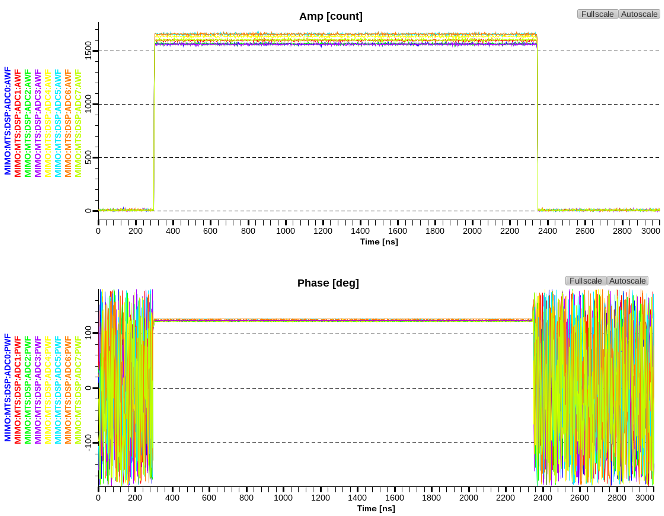}
        \caption{3GHz RF pulse loopback}
    \end{subfigure}
    \hfill
    \begin{subfigure}[b]{0.48\linewidth}
        \centering
        \includegraphics[width=\linewidth]{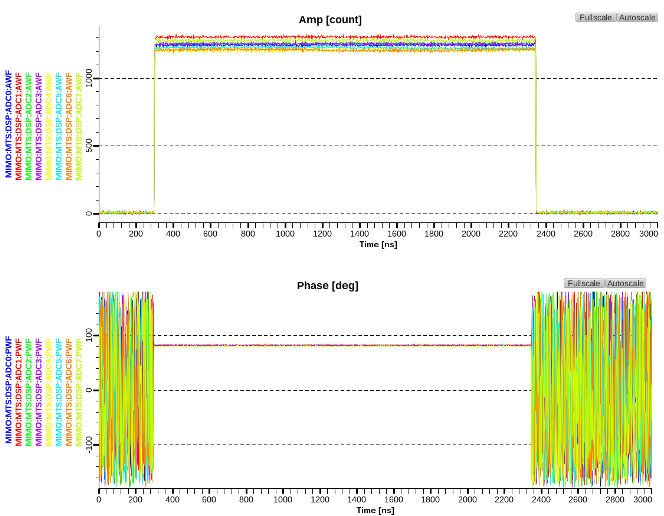}
        \caption{500MHz RF pulse loopback}
    \end{subfigure}
    \caption{EPICS Phoebus waveforms}
    \label{fig:epics_waves}
\end{figure}

 Fig ~\ref{fig:epics_waves} shows the waveforms, amplitude and phase of the looped back RF pulses at different frequencies,
  on the host Phoebus screen.

\subsection{Bench test results}

\subsubsection{Latency and mixers}
\begin{figure}[h]
    \centering
    \begin{subfigure}[b]{\linewidth}
        \centering
        \includegraphics[width=\linewidth]{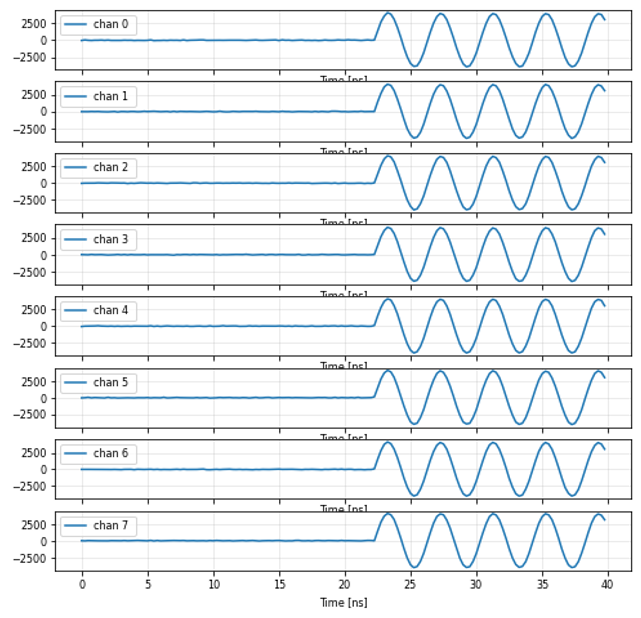}
        \caption{RAW ADC waveforms from full loopback of a pulsed RF}
    \end{subfigure}
    \begin{subfigure}[b]{0.48\linewidth}
        \centering
        \includegraphics[width=\linewidth]{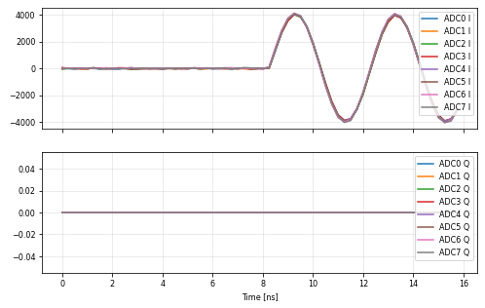}
        \caption{IQ, NCO $\theta=0^\circ$}
    \end{subfigure}
    \hfill
    \begin{subfigure}[b]{0.48\linewidth}
        \centering
        \includegraphics[width=\linewidth]{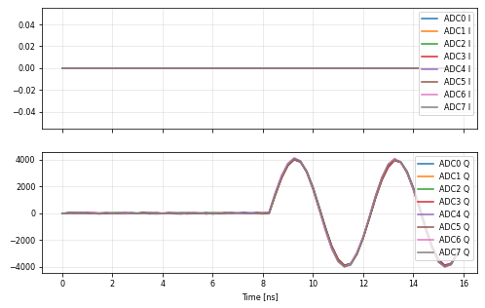}
        \caption{IQ, NCO $\theta=90^\circ$}
    \end{subfigure}
    \caption{Zoomed in at rising edge of the looped back ADC waveforms}
    \label{fig:loopback}
\end{figure}

Fig ~\ref{fig:loopback} shows measured ADC IQ waveforms when driving the DAC with a pulsed RF pulse at 250 MHz frequency.
Since all arbitrary waveform generator and waveform capturing are triggered at the same time, the total latency can be
measured by examining the rising edge of a pulsed RF waveform at the ADC.

By changing the NCO phase, IQ waveforms confirm the behavior of the RF-ADC mixers.

For the wide-band MIMO overlay, there is no ADC decimation or DAC interpolation.
A total latency of 287.25 ns is measured, and Fig ~\ref{fig:loopback} shows the rising edge.

\subsubsection{3GHz bench test}

For the ALS Linac LLRF, which has RF frequency at near 3GHz, sampled at the second Nyquist zone, we have demonstrated
a bench loop back test, with one DAC-ADC pair directly connected, and another DAC-ADC inserted with a 3GHz band-pass filter.

\begin{figure}[h]
    \centering
    \begin{subfigure}[b]{\linewidth}
        \centering
        \includegraphics[width=\linewidth]{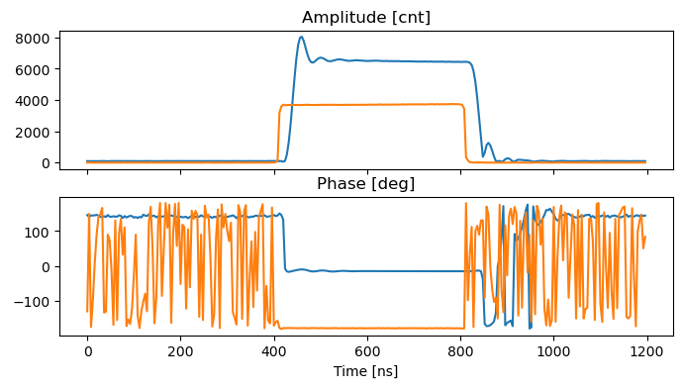}
        \caption{Base-band waves of loopback RF pulse at 3GHz.
         Orange is direct loopback, blue is through a 3GHz bandpass filter}
    \end{subfigure}
    \begin{subfigure}[b]{0.48\linewidth}
        \centering
        \includegraphics[width=\linewidth]{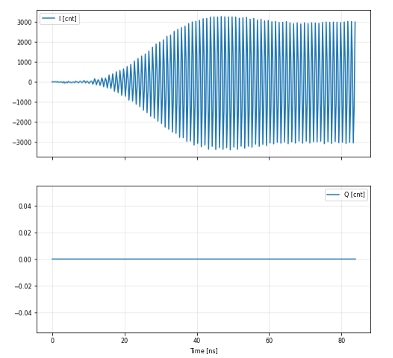}
        \caption{RAW ADC waveform}
    \end{subfigure}
    \hfill
    \begin{subfigure}[b]{0.48\linewidth}
        \centering
        \includegraphics[width=\linewidth]{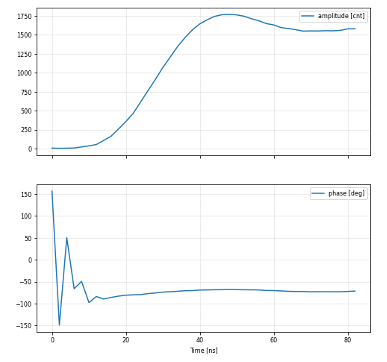}
        \caption{Base-band zoom in on filter}
    \end{subfigure}
    \caption{Bench test loopback waveforms with a 3GHz bandpass filter}
    \label{fig:linac_loopback}
\end{figure}

For this narrow-band LLRF overlay, a decimation factor of 16 and interpolation factor of 8 is added (as shown in Fig ~\ref{fig:dsp_flow}).
Fig ~\ref{fig:linac_loopback} shows the recorded ADC waveform, showing the total latency of about 400\,ns on the directly looped back channel,
and a little more on the filtered channel, representing the group delay of the band-pass filter.
Both raw IQ and baseband IQ waveforms are also shown in ~\ref{fig:linac_loopback}. This is done by changing the NCO frequency between zero and $-3000$,
using the same waveform buffer, showing the firmware flexibility and ability of LLRF signal processing.

\section{Conclusion}
\label{sec:conclusion}

We report a novel design and preliminary bench test results of a RFSoC based digital LLRF system for ALS in LBNL.
It is essential for LLRF systems to have deterministic latency, sample-to-sample data alignments, high precision
RF measurement with low channel-to-channel crosstalk.

We have evaluated the RFSoC platforms (ZCU208, LBL208) for
ALS RF frequencies, and found: a total additive RMS phase jitter on the DAC output of about 80\,fs [1Hz, 1MHz];
90dB DAC SFDR at 500MHz carrier within 200kHz span; a better than 75\,dB ADC crosstalk, -150\,dBFS/Hz ADC
 noise spectral density, and 8.9 ENOB at 500MHz input.

We have designed a modular, high isolation and low noise RF frontend, suitable for both ZCU208 and LBL208 platforms.
Programmable gain of 63 dB in RX slice and 31 dB in TX RF slice are measured at 3GHz carrier, and the linearity is 20 dB
better than the RFSoC ADC input for RX, and 7 dB better than the high power amplifier for TX.
The modular hardware design enables a enclosure design allowing
 rigid, cable-less RF connections with integrated passive thermal management.

The additive RMS phase jitter is characterized and optimized by hardware modification on the CLK104 board.
The breakdown of phase jitter contributions are analyzed, showing about 60\% is contributed by the CLK104, about 30\%
is from the RFSoC itself, and less than 1\% each is from analog RF frontend electronics.

The reported firmware and software designs allows flexible configurations of digital frequency conversions with MTS configuration at boot time.
Preliminary bench test results are shown for less than 400\,ns total latency, and the deterministic, aligned NCO phases across all channels.
Common LLRF DSP RTL designs are reused and integrated within the firmware design by open-source validation tools.
Python based development and EPICS IOC classes allows object-oriented, modular architecture for both flexibility and efficiency
 running on the RFSoC quad processors.

High power test, and non-IQ sampling schemes are planned for next steps of development.

\section*{Acknowledgment}

This work is supported by the Office of Science, Office of Basic Energy Sciences, of the U.S.
Department of Energy under Contract No. DE-AC02-05CH11231.

\bibliographystyle{IEEEtran}
\bibliography{IEEEabrv,reference}

\end{document}